\documentclass{aastex631}
\usepackage{amsmath,upgreek}

\newcommand\pennstate{Department of Astronomy \& Astrophysics and \\
Center for Exoplanets and Habitable Worlds and \\
Penn State Extraterrestrial Intelligence Center \\
525 Davey Laboratory \\
The Pennsylvania State University \\
University Park, PA, 16802, USA}

\begin{document}

\title{Potential Habitability as a Stellar Property: Effects of Model Uncertainties and Measurement Precision}

\author[0000-0003-3989-5545]{Noah W.\ Tuchow}
\affil{\pennstate}

\author[0000-0001-6160-5888]{Jason T.\ Wright}
\affil{\pennstate}

\begin{abstract}
    Knowledge of a star's evolutionary history combined with estimates of planet occurrence rates allows one to infer its relative quality as a location in the search for biosignatures, and to quantify this intuition using long-term habitability metrics. In this study, we analyse the sensitivity of the biosignature yield metrics formulated by \citet{Tuchow2020} to uncertainties in observable stellar properties and to model uncertainties. We characterize the uncertainties present in fitting a models to stellar observations by generating a stellar model with known properties and adding synthetic uncertainties in the observable properties. We scale the uncertainty in individual observables and observe the the effects on the precision of properties such as stellar mass, age, and our metrics. To determine model uncertainties we compare four well accepted stellar models using different model physics and see how they vary in terms of the values of our metrics. We determine the ability of future missions to rank target stars according to these metrics, given the current precision to which host star properties can be measured. We show that obtaining independent age constraints decreases both the model and systematic uncertainties in determining these metrics and is the most powerful way to improve assessments of the long-term habitability of planets around low mass stars. 
\end{abstract}

\section{Introduction}

In the coming decade, Astronomers strive to develop instruments with the sensitivity to search for biosignatures in the atmospheres of Earth-like planets. One of the top recommendations of the 2020 Decadal Survey on Astronomy an Astrophysics is the construction of a large IR/Optical/UV space telescope capable of directly imaging Earth-sized planets in the habitable zones of their stars, and to obtain spectra to infer their atmospheric compositions \citep{DecadalSurvey}.
In planning for such a mission, astronomers will need to select target stars to survey in order to search for planets. For direct imaging missions, only nearby bright stars would be viable candidates to host planets that are luminous enough for the instruments to obtain high enough resolution spectra. Furthermore, after planet candidates are discovered, future missions will need to prioritize which of these targets to investigate for follow-up observations.

The Decadal Survey's proposed large IR/O/UV observatory, is influenced by LUVOIR and HabEX, the two direct imaging mission concepts proposed to the decadal survey \citep{LUVOIR_final_report,HabEX_Final_report}.
These mission concepts proposed observation strategies that focused on obtaining the largest yield of spectra of habitable zone Earth-sized planets. They would select target stars for a future mission based on a variety of factors such as whether stars are known planet hosts, the amount of exozodiacal dust, and the stellar multiplicity.
While all of these properties are certainly essential in determining where one could obtain the spectra of Earth-like exoplanets, if one of the goals of a future direct imaging mission is to search for biosignatures, then target prioritization should also be based on our expectations of we expect life to emerge and develop.


In a previous study, \citet{Tuchow2020} developed a mathematical framework to determine which stars have the highest probability of hosting planets that have been in the habitable zone for an extended duration. They defined a series of \textit{biosignature yield metrics}, $B$, based on the evolution of a star's habitable zone in time, which relies on stellar effective temperature and luminosity. Thus for any stellar model which produces an evolutionary track and for any habitable zone formulation, one can compute a metric for biosignature yield in the form
\begin{equation}
        B = \iint H(a,t) \Gamma(a,R_p)\,da\,dR_p
        \label{metric_eq}
\end{equation}

The formulation of these metrics $B$ depend on two functions: $H(a,t)$ and $\Gamma(a,R_p)$. $H(a,t)$ is defined as the probability that a planet of age $t$ at a distance $a$ from its star hosts biosignatures. Lacking any empirical constraints on the emergence of biosignatures, $H$ encapsulates our prior knowledge and assumptions about the emergence of life and biosignature on planets around other stars. The function $\Gamma$ in Equation \ref{metric_eq} is the distribution of habitable planets in orbital distance. Similar to $H$, it is poorly constrained in the habitable zones of sun-like stars, though future missions aim to obtain better constraints for $\Gamma$. Therefore our knowledge of the distribution of planets in the habitable zone relies on either extrapolation to larger orbital distances, or inferences based on limited data \citep{Petigura2013,Burke2015}. Since $\Gamma$ and $H$ are not tightly constrained, there are several reasonable assumptions for both of them, comprising a family of $B$ metrics for different combinations of forms of $\Gamma$ and $H$.

The \citet{Tuchow2020} study tested various assumptions for $\Gamma$ and $H$ and determined which populations of stars were preferred by each. This study suggested that the most physical assumptions for $H$ depend on the time that planets spend in the habitable zone, which they define as the \textit{habitable duration}, $\tau$. When considering the evolution of habitable zones in time and the time spent in the habitable zone, $\tau$, one encounters the open question as to whether planets that form outside of habitable zone and enter it as their host star evolves, can actually be considered habitable. \citet{Tuchow2021} refer to the region of the habitable zone occupied by these planets as the \textit{Belatedly Habitable Zone} (BHZ) and stress that the habitability of these planets remains unknown. Since BHZ planets will make up a large fraction of planets around low mass stars and older stars, determining whether they can be habitable is a rich area of future research.  If BHZ planets are unable to support habitability, then long-term habitability would be limited to planets within the \textit{Continuously Habitable Zone} (CHZ), as defined in \citet{Tuchow2020}. 
Note that previous studies, such as \citet{Kasting1993hz} and \citet{Tuchow2020}, referred to the class of planets that form exterior to habitable zone and enter it as their host star becomes more luminous as ``cold start" planets. However, since the term ``cold start" has often been used in the context of giant planet formation, we shall instead use the term \textit{Outer BHZ Planets} to refer to them.

\citet{Tuchow2020} demonstrated how biosignature yield metrics changed as a function of stellar mass and age. Now we would like to investigate how precisions in these metrics depend on the uncertainties on stellar parameters for a typical star. 
A given stellar model can be thought of in the form $\mathbf{y} = f(\mathbf{p}, \mathbf{\theta})$. Here $\mathbf{p}$ are a set of fundamental stellar input parameters such as mass, age and metallicity, $\mathbf{\theta}$ are a variety of parameters that govern the model physics, and $\mathbf{y}$ are the model outputs. Usually stellar mass, age, and other properties in $\mathbf{p}$ are not directly observable. Instead, stellar models are typical fit to observed output properties $\mathbf{y}$, such as apparent magnitudes, effective temperatures, surface gravities, surface metallicities, and parallax measurements. For some stars, other measurements of observable properties may be available to constrain stellar models. These include but are not limited to: measurements of asteroseismic modes of oscillation, interferometric radii, dynamical masses from binaries, and gyrochronological ages inferred from stellar rotation periods.

When determining input properties $\mathbf{p}$ and derived properties, such as $B$ metrics, via fitting stellar models to observables, there are two main sources of uncertainty: systematic and model uncertainties. Systematic uncertainties arise from the fact that all observable properties have some measurement uncertainty, while model uncertainties are due to the fact that stellar models themselves are uncertain. In this study, we will analyse the sensitivity of our metrics to both systematic and model uncertainties.

The first part of our analysis will focus on systematic uncertainties.
We will observe the sensitivity of derived properties such as long-term habitability metrics, masses, and ages, to uncertainty in observable properties. We will start with fiducial stellar models with known outputs. Then we will inject synthetic uncertainties in the output parameters and see how the spread in values of the recovered input parameters and derived properties change in response. We would like to determine the effects of obtaining more precise measurements of stellar properties and to determine which properties our metrics are most sensitive to.

The second component of this study will focus on model uncertainties. While stellar structure and evolution are generally well understood for main sequence stars, stellar models often differ in terms of the physical and chemical processes they incorporate. For instance, models disagree about the chemical abundances of the solar photosphere.  This is a reflection of what is known as the Solar Abundance Problem, where newer solar abundance measurements result in a worse match to the observed speed profile obtained via asteroseismology \citep{Serenelli2009}. Stellar models also vary in terms of the equations of state, opacities, and nuclear reaction networks that are used. Most stellar models are one dimensional to make evolution calculations over billions of years computationally feasible. However, this leads to uncertainty in a variety of input physics such as their treatment of convection, convective overshoot, rotation, and element diffusion.
We will compare the outputs of four well accepted stellar models grids that vary in terms of the model physics they use. We will determine which regions of parameter space have the greatest differences between the models in terms of the calculated values for biosignature yields.

\section{Methods}
\subsection{Relative Biosignature Yield Metrics}
In this study we will make use of the relative biosignature yield metrics, $B$, defined by \citet{Tuchow2020}. These metrics are formulated in terms of assumptions about the distribution of habitable exoplanets, $\Gamma$, and the emergence of detectable biosignatures, $H$, as seen in Equation \ref{metric_eq}. Higher values for $B$ correspond to a higher likelihood of hosting planets in the habitable zone (described by $\Gamma$) that have been habitable for long enough to host biosignatures (described by $H$).

In regions of parameter space where the distribution function for exoplanets, $\Gamma$, is sufficiently well constrained, many studies show that it is well fit to either a power law or split power law \citep{Kopparapu2018,Dulz2020}.
While we don't have many empirical constraints on the distribution function of Earth-like planets in the habitable zones of sun-like stars, for this study, we will make use of metrics that assume that planet's are distributed as a power law. We will use a form of $\Gamma$ that is uniform in log semimajor axis, $\ln(a)$, which is equivalent to having a planetary distribution function in the form $C a^\beta$ with $\beta = -1$. This choice of power law has been selected for concreteness, but other realistic values of $\beta$ will lead to similar conclusions.

For our assumptions about the emergence of biosignatures, we use three physically motivated or otherwise commonly used forms for $H$, the probability that a planet with known age and semimajor axis hosts biosignatures. All of these forms of $H$ depend on the duration that a planet spends in the habitable zone, $\tau$. 

One form of $H$ that is often used, is to assume that all planets in the Continuously Habitable Zone (CHZ) have an equal chance of hosting biosignatures.  While in this study, we use the term CHZ to refer to the region of the habitable zone that remains habitable from the formation of planets to the current day, other studies often vary in terms of the formulation they use for the Continuously Habitable Zone. One of the most common formulations of the CHZ is to define it as a region occupied by planets that remains in the habitable zone longer than a given amount of time.
Values for the fixed age required for sustained habitability range in the literature from 2 Gyr to 4 Gyr \citep{Truitt2020}.
If one were to select 2 Gyr as the habitable duration required to host biosignatures, then H would have the functional form:
 \begin{equation}
 H(a,t) = 
 \begin{cases}
 \text{constant}, & \text{if } \tau(a,t) \geq 2 \text{ Gyr} \\
 0, & \text{otherwise}
 \end{cases}
 \label{fixed_age}
 \end{equation}
 Such assumptions about the emergence of biosignatures are somewhat physically motivated based on the timescales on which complex life and biosignatures developed on Earth, but having such a hard age cutoff between systems may have undesired consequences when prioritizing which stars to search for life around. 
 
 Another form of $H$ would be to assume that on a habitable planet biosignatures have a constant chance of developing, $b$, per unit time. This would have the functional form
 \begin{equation}
 H(a,t) = 1-e^{-b\tau(a,t)}   
\end{equation}
In the case where there is a low chance of biosignatures emerging per unit time, and $b$ is very small, then $H$ can be approximated as $H(a,t) \propto \tau(a,t)$. Here we actually have two possible assumptions for $H$. One could assume that BHZ planets, originating outside the habitable zone can eventually become habitable. Alternatively one could assume that only planets in the continuously habitable zone can host biosignatures. In this case, $H$ would be proportional to $\tau$ in the CHZ and zero outside of it.

With the different options for $\Gamma$ and $H$ discussed above, we have a total of three metrics that we will use in our sensitivity analyses. All these metrics have the same assumption for $\Gamma$ -- that $\Gamma$ is uniform in $\ln(a)$, but they differ in the their assumptions about $H$. First we have a metric which uses a form of $H$ which is constant for for habitable durations greater than 2 Gyr in the form of Equation \ref{fixed_age}. We will refer to this metric as the 2 Gyr metric, or \textbf{B(2Gyr)}.
We will refer to the metric with $H$ proportional to $\tau$ including BHZ planets as the BHZ metric or \textbf{B(BHZ)}. Finally, the metric with a form of $H$ that is proportional to $\tau$ in the continuously habitable zone, and zero outside of it, will be referred to as the CHZ metric or \textbf{B(CHZ)}. The values of these metrics will be normalized to correspond with the solar value of each metric. This means that a value of 1.00 for a metric corresponds to the value obtained from a model of the present day sun. 

Calculation of these metrics requires knowledge of how habitable zones change over the course of a star's lifetime. Using the temperatures and luminosities in stellar model grids discussed in the next section, one can calculate the habitable zone boundaries. In this study we will use the \citet{Ramirez2018} formulation of the classical $\text{N}_2$-$\text{CO}_2$-$\text{H}_2 \text{O}$ habitable zone. This is similar treatment of the habitable zone to that of \citet{Kopparapu2013}, but it is applicable for a wider range of stellar effective temperatures between 2,600 - 10,000 K. The inner edge of this habitable zone is defined by the \citet{Leconte2013} runaway greenhouse limit, while the outer edge is given by the maximum $\text{CO}_2$ greenhouse heating. 

\subsection{Model Grids}
\label{model_grids}
We shall make use several different stellar models in the multiple components of our sensitivity analysis. In our analysis of the sensitivity of derived model parameters to measurement uncertainties in observational parameters, we shall use a grid of MIST stellar models \citep{Dotter2016,Choi2016}. MESA Isochrones and Stellar Tracks (MIST) is a grid of stellar models computed using the MESA stellar structure and evolution code \citep{MESAPaper2011,MESAPaper2013,MESAPaper2015,MESAPaper2018,MESAPaper2019}. The MIST models we are using in this analysis include the effects of stellar rotation with a maximum rotational velocity on the zero age main sequence of $v/v_{crit} = 0.4$. 
MIST gives a grid of stellar models in mass, equivalent evolutionary phase (EEP) and metallicity. EEPs are used as a proxy for stellar ages, as stellar lifetimes vary widely over the range of masses, and if models were sampled uniformly in age, late phases of stellar evolution, where stars evolve more quickly, would have inadequate time resolution \citep{Dotter2016}. 
To infer the properties of stars between the grid points, we will use the interpolation scheme in the Isochrones python package \citep{Morton2015}.

For the sensitivity analysis for model uncertainties, we are faced with a challenge in that there are so many options to tweak for stellar model physics. Rather than varying each of the hundreds of model parameters, we will instead use the approach of \citet{Tayar2020}, comparing the results of four widely used and well established stellar models which incorporate different model physics. We will make use of their Kiauhoku python package and their model grids expressed in terms of mass, EEP, and metallicity \citep{Claytor2020, kiauhoku_code}. We will use the following model grids:
\begin{itemize}
        \item Yale Rotating Evolution Code (YREC) \citep{Pinsonneault1989, Tayar2020}
        \item MESA Isochrones and Stellar Tracks (MIST) \citep{Choi2016,Dotter2016}
        \item Dartmouth Stellar Evolution Program (DSEP) \citep{Dotter2008}
        \item Garching Stellar Evolution Code (GARSTEC) \citep{Weiss2008garstec, Serenelli2013}
\end{itemize}
For a comprehensive summary of the input physics used in each of these stellar models, see Table 1 of \citet{Tayar2020}. 



\section{Systemic/ fitting uncertainties}
\label{systemic_uncertainties}
\subsection{MCMC fitting uncertainty}
When determining the derived properties of a star, such as masses and ages, much of the uncertainty is due to uncertainties in the observed stellar properties. To obtain a stellar model for a given star, one will typically fit a model to a set of observed target properties. It is important to have several properties to fit a stellar model to, as too few constraints can lead to ambiguities in stellar parameters. For instance, if one only has measurements of luminosity and effective temperatures of stars (ie. positions on the HR Diagram), it can lead to many degeneracies between stars of different masses, ages, and metallicities \citep{GodoyRivera2021}. 
For this reason, it is useful to fit stellar models to additional measured properties such as  $\log(g)$ and surface [Fe/H] to break the degeneracy. In this section, we would like to see how improving the precision of target values for model fitting affects the precision in derived properties such as masses, ages, and our metrics for biosignature yield.

For this analysis, we will start by using fiducial stellar models of known input properties. We regard these input properties as the ``true'' properties of the stars, and we will try to recover these values by fitting to observed properties. To simulate observations, we take the model output values and add synthetic measurement uncertainties to them. 
We use the affine-invariant ensemble MCMC sampler in the emcee python package to recover stellar input properties as well as derived properties, and investigate the sensitivity of these values to the uncertainty in specific observables \citep{emcee}. We use a $\chi^2$ likelihood, comparing model outputs to observations.

In this sensitivity analysis, we consider 3 fiducial target stars (see Table \ref{target_star_table}). First we consider a star that is a solar analog with a mass of $1.0 M_\odot$, an age of 4.6 Gyr and a metallicity of [Fe/H]=0.0. Note that this model is slightly different than the sun in that the sun's metallicity isn't the same as its observed surface value. Setting [Fe/H] = 0.0 implies that the net stellar iron abundance is the same as the surface iron abundance observed for the sun. We then consider an example model for a lower mass K star. This fiducial model has a mass of $0.7 M_\odot$, an age of 5 Gyr and [Fe/H]=0.0.  Finally we consider a higher mass F star model with a mass of $1.25 M_\odot$, an age of 1 Gyr, and [Fe/H]=0.0. For all of these stars, we assume they are observed at a distance of d = 50.0 pc with a visual extinction of $A_V = 0.0$.

For each of these stars, we assume that we have measurements of the GAIA band magnitudes, parallax, $\log(g)$, $\mathrm{T_{eff}}$, and surface [Fe/H]. In a later section, we will also consider the case where independent age constraints are available. These ``measurements'' are derived from the rotating MIST model's outputs with added uncertainties typical of each measurement (see Table \ref{uncertainty_table}). For several spectroscopic quantities, we obtain typical uncertainties from the work of \citet{Brewer2016}. We adopt their typical uncertainties as 25 K in $\mathrm{T_{eff}}$, 0.010 in [M/H], and 0.028 in $\log(g)$. Note that the precision in these spectroscopic measurements may be better that those of typical field stars, but they are a good estimate for the uncertainty in the properties of a well characterized star. To obtain the uncertainties in parallax and GAIA band magnitudes we consulted the GAIA Data Release 2 \citep{GAIArelease2}.
This gives an uncertainty in parallax of 0.04 mas, while the GAIA $G$, $G_{RP}$, and $G_{BP}$ have very small uncertainties on the order of  0.1 - 1 mmag (for bright stars with $G< 13$).

Since the measurement uncertainties in band magnitudes are so small, the dominant source of uncertainty actually comes from the bolometric correction. The models which we are using utilize the MIST bolometric correction grid. These bolometric corrections are computed for the surface conditions present in the MIST models. When one applies this bolometric correction to other model grids (which we shall do in later sections), there is additional uncertainty introduced. For a different but similar bolometric correction grid, \citet{Casagrande2018} estimate the GAIA band uncertainties introduced by the bolometric correction. We will use the 0.02 mag uncertainty they obtained as a rough estimate of the uncertainty in GAIA band magnitudes caused by bolometric corrections.

\begin{figure}
    \plotone{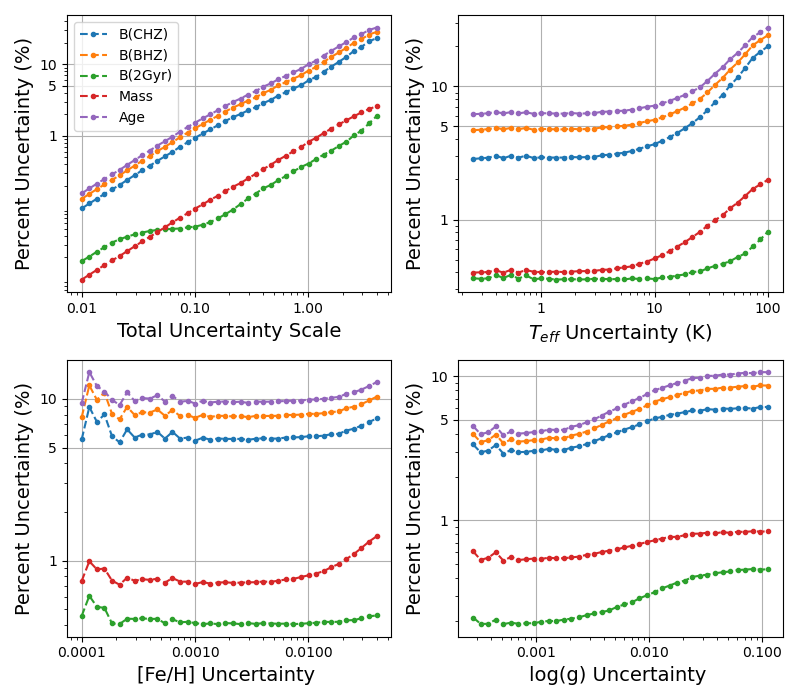}
    \caption{Percent uncertainty in stellar derived properties as a function of uncertainty in different observed stellar properties. Different panels correspond to different observables. The total uncertainty panel corresponds to scaling the uncertainty in all parameters relative to the standard values described in Section \ref{systemic_uncertainties}. These results are for the case of a 1.0 $M_\odot$, [Fe/H]=0.0 model.}
    \label{4panel_non_age}
\end{figure}

Using these standard values for the uncertainties in the target properties, we investigate the effects of scaling the uncertainty it a single observable property to higher and lower values, while the other parameters' uncertainties remain at their standard values. We vary the uncertainties in $\mathrm{T_{eff}}$, surface [Fe/H], and $\log(g)$ individually, and then scale the uncertainty in all parameters together. Using MCMC chains via the emcee python package, we observe how stellar properties such as mass, age, and the 3 biosignature yield metrics vary in response to varying the uncertainty in these target values \cite{emcee}. Figure \ref{4panel_non_age} illustrates the results of increasing precision in various target properties for a solar model. One can observe that at the standard values for the uncertainties in target values (Total uncertainty scale of 1.00 in the top left panel) different derived properties vary in terms of their typical uncertainties. Stellar ages and the habitable duration dependent metrics, B(BHZ) and B(CHZ), have the largest uncertainties between around 5 - 10\%. On the other hand stellar masses and the B(2Gyr) metric have much smaller uncertainties. 

One can see in Figure \ref{4panel_non_age} that precision in derived properties is more sensitive to certain observables that others. This figure is in the case of our example solar analog G star, but the same trends discussed below are present in the results of the other fiducial target stars.
Scaling the uncertainty in specific observables for the solar model we observe the following trends:
\begin{itemize}
    \item Decreasing uncertainties in all observables directly corresponds to a decrease in the uncertainties in derived properties. This relationship appears to be close to a power law  with similar exponents for each property.
    \item  Increasing the precision in $\mathrm{T_{eff}}$ appears to improve precision in derived properties until an uncertainty of roughly 10 K, beyond which there are diminishing returns.
    \item Increasing precision in [Fe/H] measurements does not seem to play a major role in constraining ages and our B metrics, but more imprecise measurements than the current fiducial uncertainty of 0.010 dex will lead to an increase in the uncertainties of derived properties.
    \item Measurement precision in $\log(g)$ doesn't have as strong of an effect on constraining derived properties, but decreasing uncertainty in $\log(g)$ appears to continually decrease uncertainty in ages and B metrics, down to very precise $\log(g)$ values.
\end{itemize}
Note that in this figure, the uncertainties in are scaled to unobtainably small levels. We include these small uncertainties to illustrate the effects of obtaining more precise observations, even if such precision in observations may not be attainable. If the uncertainty in properties such as [Fe/H] is unrealistically small, it starts having a chaotic effect where it dominates the goodness of fit and best fit models are found in very different regions of parameter space. Future measurements of these properties are unlikely to approach the extreme precisions where this becomes a problem, but this serves to illustrate the negative effects of having one observable measured much more precisely than the others.

\subsection{Adding Age Constraints}
We then consider the case where one has independent constraints on stellar ages. Independent age constraints are usually difficult to obtain. In some cases, methods such as gyrochronology may be able to provide independent estimates of stellar ages that can aid in the fitting of stellar models. 

\begin{figure}
    \plotone{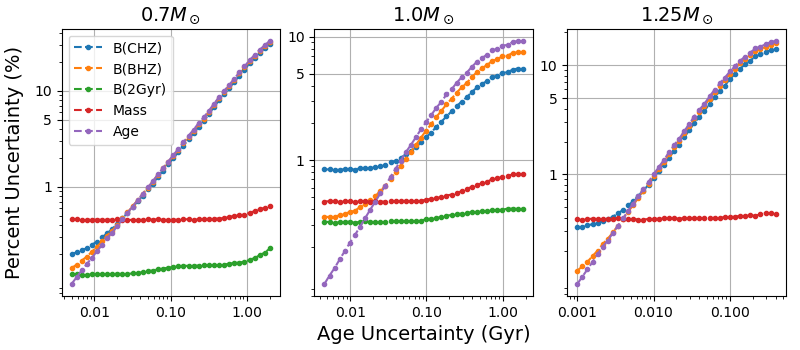}
    \caption{Percent uncertainty in stellar derived properties as a function uncertainty in stellar age constraints. The 3 panels show results for stars of different masses. The uncertainty in B(2Gyr) doesn't appear for the 1.25 $M_\odot$ star because the value for B(2Gyr) in this case is unambiguously zero.}
    \label{3panel_age}
\end{figure}

In addition to the target values described in the previous section, for the three example stars of different masses, we include an independent age constraint. We modify the precision of the age constraint and, after running MCMC chains for each value, we report the uncertainty in derived properties. 
The results of adding this age constraint can be seen in Figure \ref{3panel_age}. Obviously, adding an age constraint decreases the uncertainty in model ages for all the stars, as one can see in the figure. Less obvious is the fact that the sensitivity of B(BHZ) and B(CHZ) closely follows that of model ages, specifically in the case for the 0.7 and 1.25 $M_\odot$. Interestingly the 1.0 $M_\odot$ star appears to have lower percent uncertainties in age, B(BHZ), and B(CHZ) compared to the other stars. It is unclear why this is the case, but it may have to do with the fact that the solar analog is at a slightly later evolutionary phase than the other stars. One observes that for the solar analog, and to a lesser extent the 1.25 $M_\odot$ star, decreasing the uncertainty in age to a certain extent causes the uncertainty in B(BHZ) and B(CHZ) to saturate at a fixed value. However the age precision at which this occurs is so small as to not be reasonably obtainable. Mass and B(2Gyr) appear to be relatively insensitive to constraints in age, as they can be determined precisely without them. Note that for the case of the 1.25 $M_\odot$ star, the model ages are all less than 2 Gyr so the B(2Gyr) values are all zero with very low variability, so they don't appear on the plot.

\section{Model Uncertainties}

While systematic uncertainties in fitting stellar models to stars contribute a large portion of uncertainty in derived parameters, another major source of uncertainty comes from the fact that stellar models themselves are uncertain. In this section, we will investigate how differences between stellar models affect the calculated values for our biosignature yield metrics.

As described in Section \ref{model_grids}, we use the Kiauhoku package  to compare the outputs of four widely used stellar models with varying model physics \citep{Claytor2020,Tayar2020}. For all the model grids, we calculated the values of our B metrics for each point in mass, EEP (equivalent evolutionary phase), and metallicity space. Then we used the dataframe interpolator in the Isochrones package to interpolate our metrics between the points given in the model grids \citep{Morton2015}. 
For each of the 3 treatments of $B$, we plotted our metrics as a function of mass and EEP, and calculated the maximum difference in values between the model grids.

\begin{figure}
    \plottwo{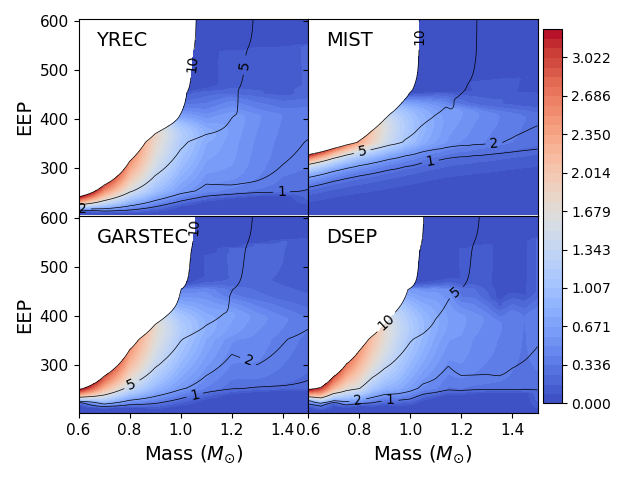}{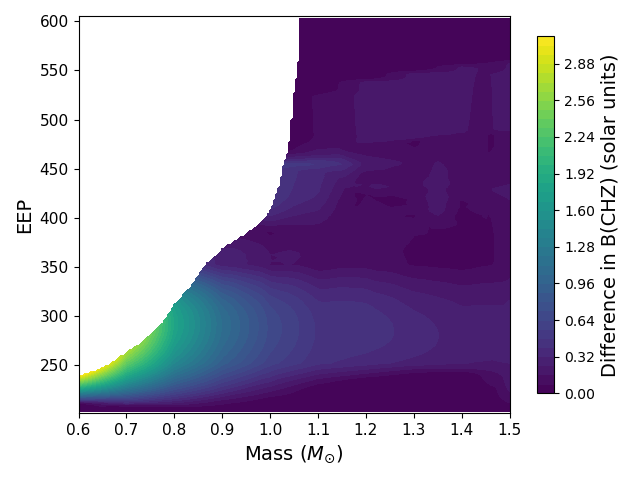}
    \caption{Difference in B(CHZ) metric between different stellar models. Left: Comparison of B(CHZ) computed for 4 stellar models grids in mass and equivalent evolutionary phase (EEP). Also included are contours of constant age in Gyr. Right: Maximum differences in B(CHZ) between stellar models. Colormaps are shown in units of the solar value of B(CHZ).}
    \label{B_CHZ_model_compare}
\end{figure}

\begin{figure}
    \plottwo{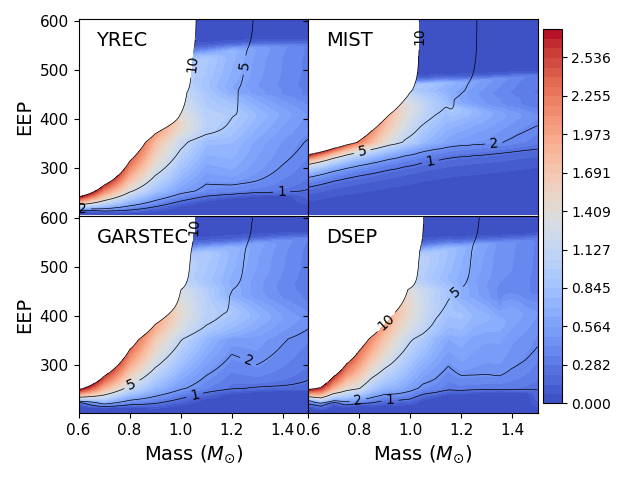}{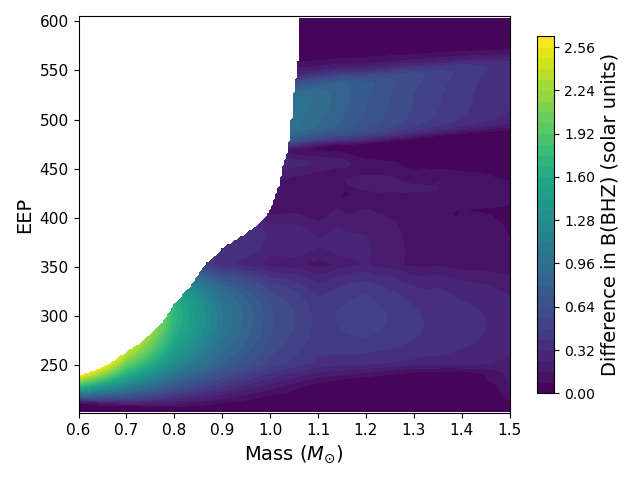}
    \caption{Difference in B(BHZ) metric between different stellar models. Left: Comparison of B(BHZ) computed for 4 stellar models grids in mass and equivalent evolutionary phase (EEP). Also included are contours of constant age in Gyr. Right: Maximum differences in B(BHZ) between stellar models. Colormaps are shown in units of the solar value of B(BHZ). }
    \label{B_CS_model_compare}
\end{figure}

\begin{figure}
    \plottwo{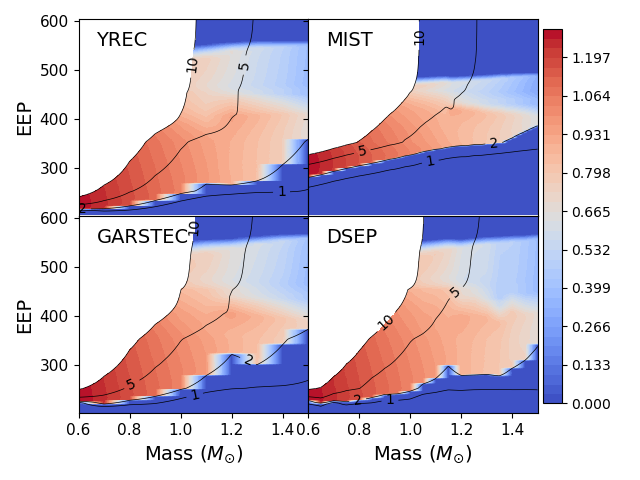}{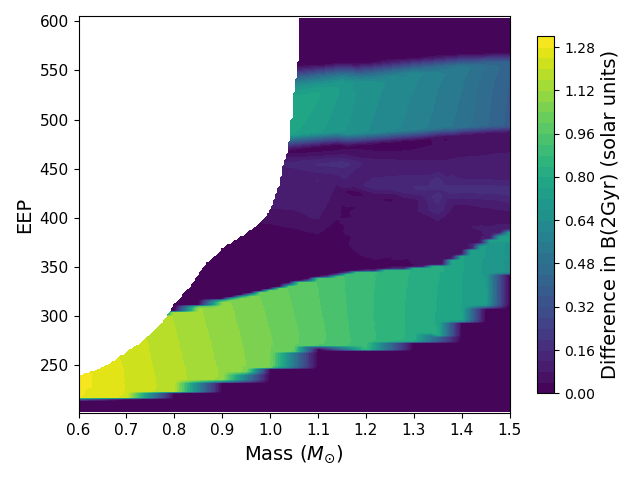}
    \caption{Difference in B(2Gyr) metric between different stellar models. Left: Comparison of B(2Gyr) computed for 4 stellar models grids in mass and equivalent evolutionary phase (EEP). Also included are contours of constant age in Gyr. Right: Maximum differences in B(2Gyr) between stellar models. Colormaps are shown in units of the solar value of B(2Gyr).}
    \label{B_2Gyr_model_compare}
\end{figure}

In Figures \ref{B_2Gyr_model_compare}, \ref{B_CHZ_model_compare}, and \ref{B_CS_model_compare}, we show the values of the B metrics for the 4 different models in the left panels, and the right panels show the maximum differences in B between models. Each of these models uses a solar metallicity of [Fe/H]=0.0. Note that these different models may disagree about what the true metal mass fraction for sun is, or what solar abundance pattern to use. However, these models agree that the solar metallicity at [Fe/H]=0.0 matches the observed surface values for the sun, even if they disagree as to what those values are. The colorbars in these figures show the values of the metrics relative to those of the current day sun.

The contours in the left panels of these figures represent stellar ages in Gyr. In these subplots, ages greater than 10 Gyr have been masked out since we are mainly concerned with nearby stars which are unlikely to have ages exceeding 10 Gyr. Furthermore, masking stars past this age gives a more informative color scale, as highly evolved low mass stars would have very high B values, but also wouldn't be found among field stars.  One can see very large differences between the age contours of different models in these plots. This represents the fact that stellar models disagree on the age of a star corresponding to a given mass and evolutionary phase. 

On the right panels of these plots are the maximum differences in B between models. For the B(CHZ) and B(BHZ) metrics, one can see that there are major differences between models during the earlier part of the main sequence, corresponding roughly to EEPs of 225 - 350. This corresponds to differences in model ages, and since these metrics are dependent on time spent in the habitable zone, they are heavily dependent on stellar ages. The mismatch between model ages is greatest for low mass stars and thus we see the largest model uncertainties for B(CHZ) and B(BHZ) in this region of parameter space. 
Additionally,  for B(CHZ) there is a slight model disagreement near the end of the main sequence at EEPs of 400 - 450. This is likely due to differences in stellar evolutionary tracks for the 4 models and disagreements about how quickly the stars evolve on the late main sequence. Since the B(CHZ) metric depends on the time spent in the continuously habitable zone, the models disagree about where B(CHZ) goes to zero, i.e. when there ceases to be a region of the habitable zone where planets have been continuously habitable from their formation. The B(BHZ) model differences plot has a similar, but more prominent secondary maximum on the subgiant branch, at EEPs around 475 - 550. The models disagree about how long regions of the habitable zones of subgiants have stayed habitable and how quickly subgiants evolve in temperature and luminosity. The B(BHZ) metric allows planets that start outside the habitable zone to be considered habitable, and depends on time spent in the habitable zone, so there is a large model discrepancy in this region.

The B(2Gyr) metric is formulated differently from the other metrics, in that it primarily depends on the width of the region that spends more that 2 Gyr in the habitable zones, and $H$ doesn't depend linearly on the habitable duration. 
This causes the plot of model differences in Figure \ref{B_2Gyr_model_compare} to appear different than those of the other metrics. In the left panel, one can see that the B(2Gyr) drops off to zero for stars with ages less than 2 Gyr, and then abruptly jumps up for ages greater than 2 Gyr. However, one can observe that the four stellar models disagree about the position of the 2 Gyr age contour. Note that in the left panels of Fig \ref{B_2Gyr_model_compare}, there is a roughness in the colormaps around the 2 Gyr contour due to the resolution of the model grids in mass and EEP spaces. The MIST model grid, with the highest resolution, doesn't have the same problems as the other grids, emphasizing that this roughness is due to resolution and not an inherent uncertainty in the models themselves. Nonetheless, resolution in model grids is a concern to take into account  when a given star falls between grid points, and it can contribute to uncertainty.
The right panel of Fig \ref{B_2Gyr_model_compare}, representing model differences in B(2Gyr), shows a large strip near the bottom of the plot representing the disagreement in the 2 Gyr contour. Furthermore there seems to be an additional region of high disagreement between models on the subgiant branch. Models disagree about when regions of the habitable zone that have remained habitable for more than 2 Gyr leave the habitable zone, and about how quickly stars evolve on the subgiant branch. 

From these results, we can see that a large portion of model uncertainties arise due to uncertainties in the ages of stars. For all of these metrics, one can see that the largest values are typically obtained for older low mass stars, but these stars also appear to have the largest model uncertainties in metric values. This is due to the fact that low mass stars evolve very slowly and it is difficult to constrain their ages as evolutionary states billions of years apart appear very similar. The B(2Gyr) metric,  which has a sharp cutoff at 2 Gyr, is particularly sensitive to these uncertainties, as models disagree about which evolutionary phase corresponds to an age of 2 Gyr. There are also model disagreements for later phases of stellar evolution, during the late main sequence and subgiant phases. While uncertainty in stellar ages may play a role in these uncertainties, the dominant source of model uncertainties for values of our metrics is likely due to differences in the evolutionary tracks of different models and disagreement about the details of the stars late main sequence and early subgiant evolution.

\section{Comparison between systematic and model uncertainties}
In a real scenario, when one fits models to stellar observations, model predictions will be subject to both model and systematic uncertainties. Using the 3 example stars described in Section \ref{systemic_uncertainties}, we will briefly discuss how the derived values for these stars vary when both these sources of uncertainty are taken into account. 

We start with the ``true" values for these stellar derived properties, based on the model outputs from the rotating MIST model used in Section \ref{systemic_uncertainties}. We would like to see whether the four different stellar models can recover the ``true" values, and whether the relative ranking of the stars are consistent and unambiguous. We will consider two cases with and without independent constraints on stellar age. For each star mass and each stellar model, we run MCMC chains and fit the models to the simulated observables.

\begin{figure}
    \plotone{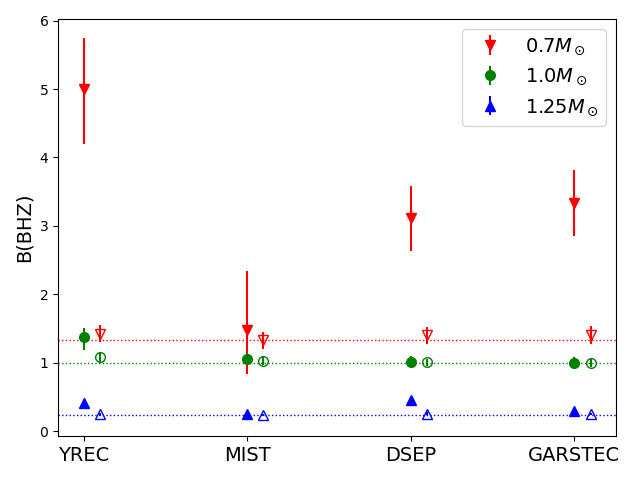}
    \caption{Comparison of recovered values for B(BHZ) for different models and different stars. Model grids are shown as columns on the X axis, while the mass of the star for which observables are being fit are shown by the colors and markers in the legend. The labels in stellar masses correspond to the set of simulated observables for each of the fiducial stellar models. Models without stellar age constraints are shown as the filled markers, while the unfilled markers show the effects of adding a $10\%$ independent age constraint. Horizontal dotted lines represent the true values of B(BHZ) from the fiducial models as a point of comparison.}
    \label{B_CS_stars_and_models}
\end{figure}

The results of these model fits can be seen in Figure \ref{B_CS_stars_and_models}. 
We plot the recovered values of the B(BHZ) metric, as this metric behaves very similarly to B(CHZ) and age, and plots of those quantities would appear almost identical. 
In this plot, the columns on the X axis represent the different stellar models and the different colored markers represent the recovered values of the B(BHZ) metric. Filled markers represent the case where one doesn't have independent age constraints, while the unfilled markers show the results if one has a $10\%$ constraint on stellar ages. The Y axis gives the value of B(BHZ) in solar units, relative to the values computed for a solar model using the rotating MIST model. Horizontal dotted lines on this plot represent the ``true" values of B(BHZ) for the fiducial model, and are shown as a point of comparison. 

Without age constraints, one can see that values for B(BHZ) appear to be quite variable between different stellar models. While for the 1.25 $M_\odot$ and solar mass cases, the values are pretty well constrained even if they may be a bit offset from the ``true" values, the B(BHZ) values for the low mass $0.7 M_\odot$ case appear to vary wildly and with large uncertainties. Only the recovered value for the non-rotating MIST model matches the "true" values, but even then the uncertainties are large enough to overlap with those of the $1.0 M_\odot$ model. Still, even if these values for B(BHZ) don't match the "true" values, one can still see that in all cases except for the MIST model, one can unambiguously rank the 0.7 $M_\odot$ star as being a better candidate for biosignatures than the 1.0 $M_\odot$ star, which in turn is a better candidate than the 1.25 $M_\odot$ star. 

The differences in B(BHZ) between different models stem from the fact that B(BHZ) is heavily dependent on a star's age, and the different models can strongly disagree on stellar ages. The disagreement is highest for low mass stars, so it makes sense that the 0.7 $M_\odot$ star has the largest model difference in B(BHZ) values. Adding a $10 \%$  independent constraint on age drastically improves the agreement between the derived B(BHZ) values and the ``true" values, especially for the case of lower mass stars. One can observe that not only is there a greater agreement between models, but there is improved precision on the derived values of B(BHZ). This demonstrates that while independent age constraints are difficult to obtain, they greatly improve our ability to constrain our values for biosignature yield metrics.

The results for the derived properties of B(CHZ) and stellar age have very similar forms to that of Figure \ref{B_CS_stars_and_models}. However, the other derived properties: mass and B(2Gyr) behave differently from these primarily age dependent quantities. The derived star masses largely match those of the fiducial models, and have relatively low uncertainties. Without age constraints there are a few small offsets from the fiducial values, but adding age constraints generally improves the agreement. For the B(2Gyr) metrics, the derived values for all the models without age constraints correspond well to the ``true" values. The uncertainties for B(2Gyr) values are generally very small, but for MCMC chains that explore the sharp boundary around 2 Gyr, such as some of the chains for the 1.25 $M_\odot$ star, the uncertainties can be quite large. Adding age constraints greatly reduces uncertainties in these values.

When both model and systematic uncertainties are included in our analysis, one can observe that stars of masses equal or greater to that of the sun can be easily ranked in terms of biosignature yield metrics. Low mass stars generally vary greatly between models and have large uncertainties, but in many cases the relative ranking between stars of different masses and ages is preserved. The incorporation of independent age constraint greatly lowers the model disagreements and uncertainties in metric values for all stars, but the effect is most prominent for the case of low mass stars.

\section{Results and Conclusions}
In this sensitivity analysis, we have shown that both model uncertainties and systematic fitting uncertainties contribute to the spread of values for our biosignature yield metrics, B, as well as for derived properties such as stellar mass and age. We looked at how several biosignature yield metrics with different models for $H(a,t)$, the probability of biosignature emergence, varied depending on the choice of stellar model and the precision of stellar measurements.

The B(2Gyr) metric treats all planets that spend more than 2 Gyr in the habitable zone as equally likely to host biosignatures. This metric therefore depends primarily on the size of the region that spends more than 2 Gyr in the habitable zone and only depends on age insofar as it affects the width of that region. B(2Gyr) has relatively small systematic uncertainties when fitting models to stars even when stellar observables are poorly constrained. However, this metric can have  large model uncertainties in some regions of parameter space, as models disagree about which stars have ages greater than or less than 2 Gyr. The sharp cutoff where B(2Gyr) drops to zero for habitable durations less than 2 Gyr makes this metric and other metrics with fixed age formulations less physical in nature and more biased by assumptions. 

The metrics B(CHZ) and B(BHZ) depend on the time spent in the continuously habitable zone and the entire habitable zone respectively. Since these metrics are proportional to the habitable duration, it makes sense that they are heavily dependent on stellar ages, and the precision in B(CHZ) and B(BHZ) depends on the precision to which the stellar ages can be determined. Without independent constraints on ages, stellar ages can be well determined for F and G spectral types, but for K stars and stars with lower masses stellar ages can be very uncertain and vary wildly between stellar models. This in turn means that B(CHZ) and B(BHZ) have the greatest discrepancy for low mass stars on the main sequence. For these metrics there are also disagreements later in a star's lifetime. B(CHZ) has high model uncertainties near the end of the main sequence, while B(BHZ) has large discrepancies between models on the sub-giant branch.

Although independent constraints on stellar ages can be very difficult to obtain, we have shown that they can play a major role in improving the precision of biosignature yield metrics. For the B(BHZ) and B(CHZ) metrics, we find that incorporating independent age constraints when fitting isochrones to stars can greatly decrease both the model and systematic uncertainties. Furthermore the precision on the age constraint applied is strongly correlated to the precision in B(BHZ) and B(CHZ) metrics. Because of the role that age constraints play in decreasing systematic and model uncertainties for these metrics we find that they make the ranking of target stars according to biosignature yield less ambiguous.

In preparation for future space-based direct imaging missions, it makes sense to place emphasis on obtaining independent stellar age constraints. This would allow target stars to be prioritized according to their long-term habitability and would allow future missions to test hypotheses involving stellar ages, such as the planetary age-oxygen correlation proposed as a statistical biosignature \citep{Bixel2020}. Ongoing missions such as TESS are able to obtain stellar rotation periods which can be used to obtain age constraints via gyrochronology, as well as asteroseismic modes of oscillation which can further constrain models. With more independent age constraints for the population of nearby stars, we will be able to better assess the long-term habitability of lower mass stars, and determine the best candidates in a search for biosignatures. 

Even without age constraints, the long-term habitability metrics which we have tested in this analysis will be useful for prioritizing direct imaging observations. While one may have difficulty determining the relative ranking of individual low mass stars according to these metrics, ultimately what is most important is determining which populations have higher metric values relative to each other. The specific values of a given metric are determined probabilistically, based on planet occurrence rates and the probability of biosignature emergence. This means that the exact value of a metric for a specific star is less informative than the relative values between stars and for populations. Future missions will be able to select optimal target lists according to these long-term habitability metrics regardless of whether we have the precision to determine an unambiguous ranking of specific stars according to these metrics. By using these metrics as a factor for target prioritization, future missions can ensure that they have the highest chance of observing biosignatures in planetary atmospheres.

    
    

\begin{table}
    \centering
    \begin{tabular}{|l|l|l|l|}
        \hline
         Model & K star & G star & F star \\
         \hline
         Mass ($M_\odot$) & 0.7  & 1.0 & 1.25 \\
         Age (Gyr)        &  5.0   & 4.6 & 1.0     \\
         $\textrm{[Fe/H]}$ & 0.0 & 0.0 & 0.0 \\
         distance (pc) & 50.0 & 50.0 & 50.0  \\
         $A_V$         & 0.0 & 0.0 & 0.0 \\
         \hline
    \end{tabular}
    \caption{True values for fiducial target stars}
    \label{target_star_table}
\end{table}

\begin{table}
    \centering
    \begin{tabular}{|l|l|}
        \hline
        Property  & Standard Uncertainty \\
        \hline
        $\mathrm{T_{eff}}$ & 25 K \\
        $\textrm{[Fe/H]}_s$ & 0.010 \\
        $\log(g)$ & 0.028 \\
        parallax & 0.04 mas \\
        GAIA bands & 0.02 mag \\
        \hline
    \end{tabular}
    \caption{Typical uncertainties in stellar observable properties}
    \label{uncertainty_table}
\end{table}

\section*{Acknowledgements}
NWT and JTW acknowledge James Kasting and Ravi Kopparapu for their assistance and guidance on this project. This work has made use of data from the European Space Agency (ESA) mission
{\it Gaia} (\url{https://www.cosmos.esa.int/gaia}), processed by the {\it Gaia}
Data Processing and Analysis Consortium (DPAC,
\url{https://www.cosmos.esa.int/web/gaia/dpac/consortium}). Funding for the DPAC
has been provided by national institutions, in particular the institutions
participating in the {\it Gaia} Multilateral Agreement.

The computations for this study made use of the following python packages: Isochrones (\url{https://github.com/timothydmorton/isochrones}), Kiauhoku (\url{https://github.com/zclaytor/kiauhoku}), and emcee (\url{https://github.com/dfm/emcee}).

The Center for Exoplanets and Habitable Worlds and the Penn State Extraterrestrial Intelligence Centers are supported by the Pennsylvania State University and the Eberly College of Science.

\bibliographystyle{aasjournal}
\bibliography{sources}
\end{document}